\documentclass[manuscript]{acmart}
\usepackage{arydshln}
\usepackage{amsmath}
\usepackage{multirow}
\usepackage{xcolor,colortbl}

\usepackage[show]{chato-notes}

\setcopyright{acmcopyright}
\acmDOI{10.1145/1122445.1122456}

\acmConference[RecSys '20]{RecSys '20:  The 14th ACM Recommender Systems Conference}{Sept 22--26, 2020}{online}
\acmPrice{15.00}
\acmISBN{978-1-4503-9999-9/18/06}

\usepackage{subcaption}

\newcommand{\ric}[1]{\textcolor{black}{#1}}
\newcommand{\zm}[1]{\textcolor{black}{#1}}

\newcommand{\zqc}[1]{\textcolor{black}{#1}}
\newcommand{\ricr}[1]{\textcolor{black}{#1}}

\usepackage{marginnote}

\begin{document}
%\title{Exploring Data Splitting Strategies for \protect\\ Neural Recommendation Models}
\title{Exploring Data Splitting Strategies for the Evaluation\protect\\  of Recommendation Models}

\author{Zaiqiao Meng}
\author{Richard McCreadie}
\author{Craig Macdonald}
\author{Iadh Ounis}
\email{(firstname.lastname)@glasgow.ac.uk}
\affiliation{%
	\institution{University of Glasgow}
}

\begin{abstract}
	Effective methodologies for evaluating recommender systems are critical, so that \zqc{different} systems can be compared in a sound manner. A commonly overlooked aspect of \zqc{evaluating recommender systems} is the selection of the data splitting strategy. In this paper, we both show that there is no standard splitting strategy and that the selection of splitting strategy can have a strong impact on the ranking of recommender systems. In particular, we perform experiments comparing three common splitting strategies, examining their impact over seven state-of-the-art recommendation models \zqc{on} two datasets. Our results demonstrate that the splitting strategy employed is an important confounding variable that can markedly alter the ranking of state-of-the-art \zqc{recommender} systems, making much of the currently published literature non-comparable, even when the same datasets and metrics are used.
\end{abstract}
\maketitle

\section{Introduction}
\ric{Recommender systems \zqc{(RecSys)} have been subject to extensive research examining how to most effectively find items of interest that a user would like to buy or consume within large datasets. Recommendation spans a range of domain-specific sub-tasks (such as grocery recommendation~\cite{wan2018representing} and venue recommendation~\cite{manotumruksa2018contextual}) and different scenarios (such as session-based recommendation~\cite{zhang2019deep} and sequential recommendation~\cite{quadrana2018sequence}). Many approaches have been proposed to solve these tasks over the last two decades, among which neural network-based recommendation models are currently very popular, due to their high effectiveness and adaptability to different sub-tasks and scenarios~\cite{zhang2019deep}. As the recommender systems field matures, advances in performance naturally become more incremental, leading to smaller increases in model effectiveness. This places more strain on the evaluation methodology's ability to distinguish between systems with similar performance, as researchers and practitioners chase ever smaller performance gains.}
\enlargethispage{1\baselineskip}

\looseness -1 \ric{With the current influx of very similar neural network-based recommendation models being published, there needs to be increased emphasis placed on eliminating confounding factors that can lead to uncertainty during evaluation, otherwise it will be impossible to confidently determine whether gains are truly being made. In the Information Retrieval (IR) domain, standardization efforts such as TREC, and other evaluation initiatives like NTCIR, CLEF and FIRE laid down guidelines on what constitutes a sound evaluation methodology in that domain. However, standardization efforts in the recommender systems domain appear to have been less successful, with most current research papers reporting a wide-range of distinct combinations of datasets, \zqc{metrics,} baselines and data splitting strategies, which makes it difficult to measure progress in the field~\cite{zhang2019deep,dacrema2019we,rendle2020neural}.}

\looseness -1 \ric{Standardization of datasets and baselines within the \ricr{RecSys community} is an on-going process. \ricr{In particular}\zqc{, while recent works~\cite{dacrema2019we,rendle2019difficulty, rendle2020neural} \ricr{tend to share similar baseline models (e.g. some variant on BPR~\cite{rendle2009bpr}) and in some cases may share datasets, there are no commonly agreed-upon standards for important aspects that can impact performance such as data preparation. Indeed, a recent study}~\cite{rendle2019difficulty} \ricr{found that suitably tuned baselines could in some cases match or out-perform state-of-the-art approaches, highlighting the importance of} hyper-parameter tuning and standardized benchmarks~\cite{dacrema2019we,rendle2019difficulty, rendle2020neural} \ricr{to enable fair comparisons and reproducibility.}}  However, \ricr{beyond these known issues, one factor} that is often overlooked \ricr{(and typically is not detailed sufficiently in prior works to be reproducible)} is the \emph{data splitting strategy} employed. This is how a \zqc{recommendation} dataset is split into training\zqc{, validation and testing} sets.  In the IR domain, this split is usually explicitly defined by the test collection (i.e. training and test query sets). However, there is often no equivalent guidance in \zqc{RecSys} scenarios, leading to a wide range of strategies for dividing any particular dataset being employed and reported~\cite{shani2011evaluating,zhang2019deep,quadrana2018sequence,said2014compare}. Hence, it is natural to ask `does the data splitting strategy matter?', because if it does, much of the recently published work is not comparable, even when performances are reported under the same dataset and metrics. As such, in this paper, we make an analysis of data splitting strategies for next-item/basket recommendation tasks, with the aim of answering this question.}

\ric{Indeed, when analysing the literature, we found many inconsistencies in terms of the rankings of different state-of-the-art neural recommendation models~\cite{dacrema2019we}. Furthermore, some prior works~\cite{zhang2019deep,quadrana2018sequence} have indicated that an arbitrary choice \ricr{of} dataset split removes (temporal) recommendation signals that some models aim to leverage. We hypothesize that some of the inconsistencies observed may be caused by particular models being sensitive to the data splitting strateg\ricr{y} employed. To validate our hypothesis, we collect and analyze the commonly used data splitting strategies among the state-of-the-art recommendation approaches (particularly the recent neural recommendation approaches) and conduct a comprehensive comparison of algorithms' performance under these strategies.}

\enlargethispage{1\baselineskip}

\looseness -1 \ric{The contribution of this work is twofold: (1) we report an analysis of recent recommendation literature to illustrate the large variance of data splitting strategies currently being employed; (2) we make a comprehensive analysis of the performance for several state-of-the-art recommendation models over three different data \zqc{splitting} strategies to evaluate the impact of those strategies. \zqc{\ricr{Indeed, o}ur analysis highlights \ricr{the often ignored limitation} that the \zqc{leave one last and the temporal user split} strategies \ricr{have, namely that they} `leak' evidence from future interactions into the model during training.} Furthermore, we demonstrate that the\ricr{se} different data splitting strategies strongly impact the ranking of systems under the same dataset and metrics - confirming that the data splitting strategy is a confounding variable that needs to be standardized. We also provide best practice recommendations for future researchers based on our analysis.}

% JTM  NIPS 2019 user-based split
\section{Data Splitting Strategies in Recommendation Models}\label{sec:strategies}
% There basically three types of data split
% Different method use different data split, and the results are different even using the same dataset model.
% Proposed our augment
\begin{table}[tb]
	\caption{\ricr{Overview of data splitting strategies reported in the literature, as well as the dataset(s) those papers use.}}
	\vspace{-3mm}
	\label{tab:models}
	\resizebox{0.80\textwidth}{!}{
		\begin{tabular}{cccccccp{1.6cm}}
			\toprule
			\multirow{2}{*}{\textbf{Model}} & \multicolumn{2}{c}{\textbf{Leave One Last}} & \multicolumn{2}{c}{\textbf{Temporal Split}} &  \multirow{2}{*}{\textbf{Random Split}}& \multirow{2}{*}{\textbf{User Split}} &\multirow{2}{*}{\textbf{Used Datasets}} \\
			\cmidrule{2-5}
			&  Item & Basket/Session & User-based & Global \\
			\midrule
			BPR~\cite{rendle2009bpr} (2009) &  $\times$ & $\times$ & $\times$ & $\times$ & $\surd$ &$\times$ & N \\
			FPMC~\cite{rendle2010factorizing} (2010) &   $\times$ &  $\surd$ & $\times$ & $\times$ &$\times$ & $\times$ & -\\
			NeuMF~\cite{he2017neural} (2017) &  $\surd$ & $\times$ & $\times$ & $\times$ & $\times$ &$\times$ & M1, P \\
			VAECF~\cite{liang2018variational} ((2018)) &  $\times$ & $\times$ &  $\surd$ & $\times$  & $\times$  & $\surd$ & M2, N\\
			Triple2vec~\cite{wan2018representing} (2018) &  $\times$ & $\surd$ &  $\times$ & $\times$ & $\times$ & $\times$ & I, D \\
			SARRec~\cite{kang2018self} (2018) & $\surd$  & $\times$ & $\times$ & $\times$  & $\times$ & $\times$  & A, M1 \\
            CTRec ~\cite{bai2019ctrec} (2019) &   $\surd$& $\surd$ & $\surd$ & $\times$  & $\times$ & $\times$ & T, A\\
			SVAE ~\cite{sachdeva2019sequential} (2019) &   $\times$& $\times$ & $\surd$ & $\times$  & $\times$ & $\surd$ & M1, N\\
			BERT4Rec ~\cite{sun2019bert4rec}(2019) & $\surd$  & $\times$  & $\times$  & $\times$ &  $\times$ & $\times$ & A, M1, M2\\
			NGCF~\cite{wang2019neural} (2019) &  $\times$ & $\times$ & $\times$  & $\times$ & $\surd$ & $\times$ & A, G, Y\\
			VBCAR~\cite{meng2019variational} (2019) & $\times$ &  $\times$& $\times$ & $\surd$  &$\times$ & $\times$ & I\\
			KGAT~\cite{wang2019kgat} (2019) & $\times$ &  $\times$& $\surd$ &  $\times$ & $\times$ &$\times$ & A, Y\\
            Set2Set~\cite{hu2019sets2sets} (2019) & $\times$ &  $\times$& $\surd$ &  $\times$ &$\times$ & $\times$ & T, D\\
			DCRL~\cite{xiao2019dynamic} (2019) & $\times$  & $\times$ & $\times$ & $\surd$ & $\times$ &$\times$ & M2, G \\
			TiSASRec~\cite{li2020time} (2020) & $\surd$  & $\times$ & $\times$ & $\times$ & $\times$ &$\times$ & M1, A \\
            JSR~\cite{zamani2020learning} (2020) & $\surd$  & $\times$ & $\times$ & $\times$ & $\times$ & $\times$ & M2, A \\
            HashGNN~\cite{tan2020learning} (2020) & $\times$  & $\times$ & $\times$ & $\times$ & $\surd$ & $\times$ & M2, A \\
             \bottomrule
			\multicolumn{7}{c}{M1: Movielens-1M, M2: Movielens-20M, T: Tafeng, D: Dunnhumby}\\
			\multicolumn{7}{c}{G: Gowalla, I: Instacart N: Netflix, A: Amazon, Y: Yelp, P: Pinterest}
		\end{tabular}
	}
	\vspace{-3mm}
\end{table}
\ric{Among the different recommendation system evaluation approaches available, ``offline'' evaluation using historical item ratings or implicit item feedback are by far the most common. As this method relies on a dataset of prior explicit or implicit interactions and current models are based on supervised learning, the dataset needs to be split into training, validation and \zm{testing} sets. We summarize the \zqc{four} main data splitting (partition) strategies from the literature below: }

\vspace{1mm}
\noindent \ric{\textbf{Leave One Last}: As its name suggests, \zqc{leave one last data splitting extracts the final transaction per user for testing, where the second last \ricr{transaction} per use\ricr{r} is normally used as validation and the remaining \ricr{transactions} can be used for training.} \ricr{There are two common Leave One Last strategies employed based on the type of transaction involved:}}

\vspace{-1mm}
\begin{itemize}
\item \emph{Leave One Last \textbf{Item}:} \ricr{Under Leave One Last Item, a transaction corresponds to one} $\langle user,item \rangle$ pair per-user. This is one of the most commonly reported strategies in the literature for item-based recommendation tasks. For example, NeuMF~\cite{he2017neural}, CTRec~\cite{bai2019ctrec} and \zqc{JSR~\cite{zamani2020learning}} models were reported using this data splitting strategy.
\item \emph{Leave One Last \textbf{Basket/Session}:} Under Leave One Basket/Session Out a transaction corresponds to a basket or session (i.e. a $\langle user,[item_1,...,item_k]\rangle$ tuple) for \ricr{each} user. \ricr{This strategy is commonly reported }in \ricr{scenarios} where an interaction represents a set of items bought together (e.g. in grocery recommendation) \ricr{where} the last basket per user in the dataset is used for testing (e.g. FPMC~\cite{rendle2010factorizing}, Triple2vec~\cite{wan2018representing} and CTRec~\cite{bai2019ctrec}).
\end{itemize}

\begin{figure}[tb]
	\centering
	\includegraphics[width = 75mm]{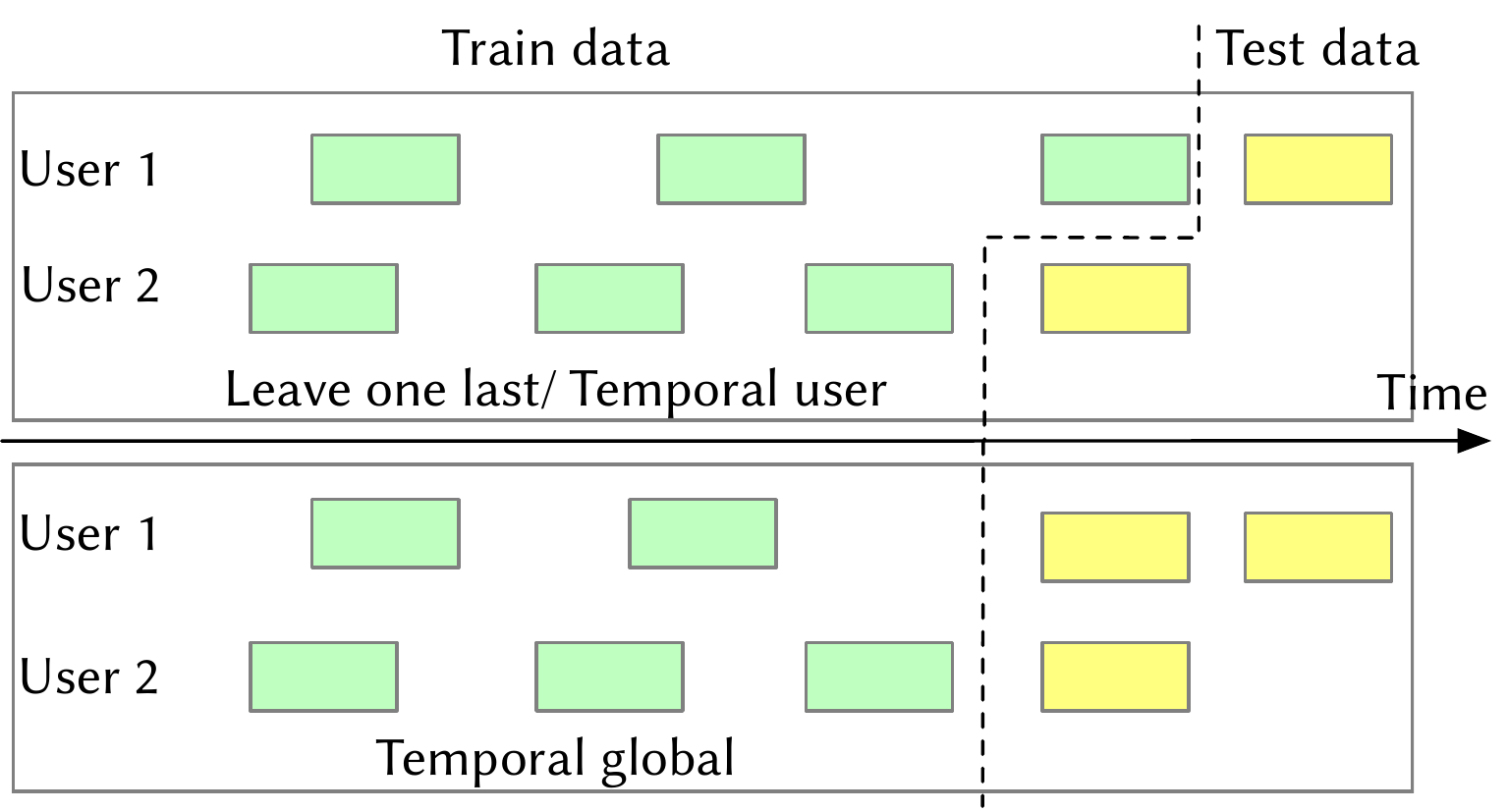}
	\vspace{-3mm}
	\caption{\label{fig:types}Temporal global split v.s. Temporal user split/Leave one last}
	\vspace{-3mm}
\end{figure}

\looseness -1 {Among our analysed papers, \zqc{leave one last} data splitting (either item or basket) was the most popular (\zqc{8 out of 17}).} \zqc{The advantage of th\ricr{ese} data split\ricr{ting} strategies is that they maximize the \ricr{number of transactions in the} dataset \ricr{that can be used} for training. \ricr{On the other hand, as} only the last \ricr{transaction} per user is \ricr{leveraged for} testing\ricr{,} test performance may not reflect \ricr{the overall recommendation effectiveness for a user over time. This also impacts training, as validation on such a small sample may not be sufficiently robust to enable consistent convergence into an effective and generalizable model.}} Moreover, the leave-one-last split strategies may cause the `leaking' phenomenon that feature interaction `leaking' into the model during the training. \zm{This `leaking' phenomenon may cause some undesirable training, for example, the model learns about the popularity of an item BEFORE it becomes popular.}  Figure~\ref{fig:types} illustrates this effect. 

\enlargethispage{1\baselineskip}

\vspace{1mm}
\looseness -1 \noindent \ric{\textbf{Temporal User/Global Split}: The temporal split strategy is another commonly used evaluation approach that splits the historical interactions/baskets by percentage based on the interaction timestamps (e.g. the last 20\% of interactions are used for testing). However, there are two variations of this strategy, which we denote temporal user and temporal global:}

\begin{itemize}
\item \emph{Temporal \textbf{User}:} Temporal user-based splitting is very similar to the \zqc{leave one last} strategy, but with the distinction that a percentage of the last interactions/baskets of each user are reserved for testing, rather than just one. Models such as VAECF~\cite{liang2018variational}, SVAE~\cite{sachdeva2019sequential} and NGCF~\cite{wang2019neural}) were originally evaluated under this strategy. It is important to note that while temporal user-based splitting does consider the \zqc{global} interaction timestamps, it is still not a realistic scenario since the train/test boundary can vary considerably between users, resulting in evidence about future interactions `leaking' into the model during training. 
\item \emph{Temporal \textbf{Global}:} On the other hand, temporal global splitting defines a fixed time-point that is shared across all users, where any interactions after that point are used for testing. VBCAR~\cite{meng2019variational} and DCRL~\cite{xiao2019dynamic} use this strategy and earlier work considered this to be the most strict and realistic setting~\cite{campos2011towards}. \zqc{However, one limitation of the temporal global splitting is that \ricr{after calculating} the intersection between the training and testing sets \ricr{(as users/items may no longer exist in both)}, the total number of users and items retained is \ricr{much smaller than under the Leave One Last strategies} (see Table~\ref{tab:dataset} for an example on \ricr{the} Tafeng dataset)\ricr{, meaning fewer transactions are available for training/validation/testing.}}
\end{itemize}

\looseness -1 \noindent \textbf{Random Split}:  \ricr{As the name suggests, random splitting randomly selects the training/test boundary per-user}~\cite{rendle2009bpr,tan2020learning,wang2019neural}.  Early recommender systems were evaluated using a leave one variant of this scheme~\cite{rendle2009bpr}, where only one random item per user is selected for testing. However, this scheme has been gradually abandoned in favour of using the last (in time) interaction (i.e. Leave One Last Item)  for each user. One limitation of random splitting strategies is that \ricr{they are not reproducible unless the data splits used are released by the author(s).}

\vspace{1mm}
\looseness -1 \noindent \ric{\textbf{User Split}: The user split strategy is another less common evaluation approach that splits the dataset by user rather than by interaction. In this case, particular users \ricr{(and hence their transactions)} are reserved for training, while a \ricr{different user set (and their transactions)} are used for testing.  Few works use this strategy, as it requires that \ricr{the underlying models have the capability to recommend items for new (cold-start) users, which many approaches do not support.}} It is also notable that some papers (e.g. VAECF~\cite{liang2018variational} and SAVE~\cite{quadrana2018sequence}) that use this strategy, still split the interaction history of the training users into \emph{fold-in} and \emph{fold-out} sets, \ricr{such that users with partial histories are included during training.} These works suffer from the same issue of ``future data'' leaking into the model during training as with the temporal user-based strategy.

\enlargethispage{1\baselineskip}

\vspace{1mm}
\ric{To provide an overview of where these strategies are being used, we analyze \zqc{seventeen} prior papers that propose and evaluate recommendation models (focusing on recent neural network approaches) and categorize them by the data splitting strategies employed. Table~\ref{tab:models} summarizes what strategies are employed by each prior work. As we can see from Table~\ref{tab:models}, there is little in the way of consistency in terms of the data splitting strategy used/reported, even in cases where two works use the same datasets. For example, the VAECF~\cite{liang2018variational} and TiSASRec~\cite{li2020time} models use the same Movielens-1M dataset, but are tested under \zqc{leave one last} and temporal splitting strategies respectively. Furthermore, we can see from Table~\ref{tab:models} that very few (\zqc{2 out of 17}) models are being evaluated using what is considered to be the most realistic splitting strategy~\cite{campos2011towards} - temporal global splitting.}

\section{Evaluation Methodology}

\begin{comment}
\begin{table}[tb]
	\caption{Statistics of the datasets used in our experiments.}
	\vspace{-3mm}
	\label{tab:dataset}
	\begin{tabular}{@{}l*{4}c@{}}
		\toprule
		\textbf{Dataset} &  \textbf{\#Users} & \textbf{\#Items} & \textbf{\#Baskets} & \textbf{\#Interactions}
		\tabularnewline
		\cmidrule{2-5}
		\textbf{Tafeng} & 9,238 & 7,973 & 77,202 & 464,118
		\tabularnewline
		\textbf{Dunnhumby} & 2,497 & 61,600 & 113, 831 & 1,048,575 \\
		\bottomrule
	\end{tabular}
    \vspace{-0.5em}
\end{table}
\end{comment}

\ric{Having shown that prior works report performance under a wide range of data splitting strategies, we next answer our primary research question: `Does the data splitting strategy matter?' In the remainder of this section we describe our experimental setup for answering this question.}

\vspace{1mm}
\noindent \textbf{Data Split}:
\ric{We experiment with three \ricr{of the most popular} data splitting strategies discussed above, namely:  \emph{leave one last item}; \emph{leave one last basket}; and \emph{global temporal split} strategies. User-based temporal split produces near-identical splits to leave one last item, and hence we exclude it to save space. Meanwhile, we omit the user split scheme since it is both rarely used and mandates a very different evaluation pipeline~\cite{quadrana2018sequence}.}

\begin{table}[tb]
	\caption{Statistics of the datasets used in our experiments. \ricr{Interactions are reported by training/validation/test sets post sampling.}}
	\vspace{-3mm}
	\label{tab:dataset}
	\resizebox{120mm}{!}{
	\begin{tabular}{@{}lc*{4}c@{}}
		\toprule
		\textbf{Dataset} & \textbf{Data split} & \textbf{\#Users} & \textbf{\#Items} & \textbf{\#Baskets} & \textbf{\#Interactions}
		\tabularnewline
		\cmidrule{1-6}
		\multirow{3}{*}{\textbf{Tafeng}}
		& leave one item & 9,238 & 7,857 & - & 444,207 / 9,238/ 9,238 \\
		& leave one basket & 9,238 & 7,857 & 58,654 & 346,378 / 58,076 / 58,229 \\
		& global temporal & 1,997 & 2,017 & 20,190 & 83,374 / 26,408 / 18,107 \\
		\midrule
		\multirow{3}{*}{\textbf{Dunnhumby}}
		& leave one item & 2,492 & 23,404 & - & 2,379,184 / 2,492 / 2,492 \\
		& leave one basket & 2,486 & 23,404 & 261,976 & 2,330,466 / 26,610 / 26,951 \\
		& global temporal & 2,162 & 25,393 & 84,128 & 715,007 / 156,476 / 169,578\\
		\bottomrule
	\end{tabular}}
	\vspace{-3mm}
\end{table}

\noindent \textbf{Datasets}:
\looseness -1 \ric{We conduct experiments on two real-world grocery transaction datasets, namely the \emph{Tafeng}\footnote{\url{http://www.bigdatalab.ac.cn/benchmark/bm/dd?data=Ta-Feng}} and \emph{Dunnhumby}\footnote{\url{http://www.dunnhumby.com/careers/engineering/sourcefiles}} datasets, \zm{which both contain} the needed information (i.e. interactions, baskets and timestamps) for the three data splitting strategies we examine~\cite{wan2018representing}. For the leave one last item/basket data splitting strategies, we \zm{first} filter items that \ricr{were} purchased less than 10 times, \zm{then} use the most recent item/basket for testing, the second recent item/basket for validation and the remaining items/baskets for training. For the global temporal split, any user that has purchased less than 30 items and/or has less than 10 baskets is filtered out, and any item that was purchased less than 20 times is removed, following \cite{meng2019variational}. Then, we split all the baskets for each of the datasets into training (80\%) and testing (20\%) \zm{subsets} based on time order, where the last 20\% of the training \zm{subset} is used for validation. Note that under the global temporal split strategy, the number of test users is further reduced, since only users that have an item/basket after the global temporal boundary are used (this particularly impacts the Tafeng dataset). We report the statistics of each dataset under each splitting strategy in Table ~\ref{tab:dataset}.}

\vspace{1mm}
\noindent \textbf{Testing Models}:
To determine the impact of the splitting strategy, we experiment with a set of seven recommenders from the literature. First, we include two classical models (i.e. NMF~\cite{lee2001algorithms} and BPR~\cite{rendle2009bpr}). Second, we select three state-of-the-art neural item recommendation models that have been shown to be effective (NeuMF ~\cite{he2017neural}, VAECF~\cite{liang2018variational} and NGCF~\cite{wang2019neural}). Finally, we include two state-of-the-art neural grocery recommendation models (Triple2vec~\cite{wan2018representing} and VBCAR~\cite{meng2019variational}). All models support all splitting strategies. For our first experiment using all seven models, for algorithms that have hyper-parameters, we use the recommended values from the original works (either the associated paper or source code). For the second experiment, we \ricr{tune} NeuMF ~\cite{he2017neural}, Triple2vec~\cite{wan2018representing} and VBCAR~\cite{meng2019variational} using different hyper parameter settings (embedding size, learning rate, activator, optimiser and alpha values).

\enlargethispage{1\baselineskip}
\vspace{1mm}
\noindent \textbf{Evaluation Metrics}: \ric{The two commonly used ranking metrics, NDCG@10 and Recall@10, are used to evaluate the performance for each model. To quantify differences in pairs of model rankings for \ricr{the} different splitting strategies we also report ranking correlation via \textbf{Kendall's $\tau$}.}

\section{Results}

\looseness -1 In Section~\ref{sec:strategies} we demonstrated that prior works in item recommendation use very different data splitting strategies, even in cases where the dataset is the same. This is problematic, since even if the dataset and metrics reported are the same in two different papers, the performance numbers may not be comparable due to the confounding variable that is the splitting strategy. Hence, in this section, we investigate what impact the splitting strategy has on a range of classical and state-of-the-art recommendation models. In particular, we compare the ranking of systems produced under three commonly used splitting strategies: leave one last item, leave one last basket and temporal \zm{global split}. If the ranking of systems significantly differ between splitting strategies, then this serves to demonstrate that much of the recent work in the recommendation space is not comparable, and hence there is a growing need for evaluation standardization.

\looseness -1 Table~\ref{tab:swaps} reports the ranking of 7 recommendation models from the literature under four scenarios (the combination of two datasets and two evaluation metrics) for each of the three data splitting strategies. The rows are sorted by performance under leave one last (item) splitting, where the up/down arrows indicate relative rank position swaps and the number in brackets indicates the number of ranks moved. As we can see from Table~\ref{tab:swaps}, under all four scenarios, rank swaps are observed between systems. For example, for the \zqc{Dunnhumby} dataset under Recall@10, the worst model under leave one last item (NMF) is ranked three places higher under leave one last basket, passing BPR, VAECF and NGCF. Indeed, we observe swaps occurring for all pairs of splitting strategies, and more worryingly, these swaps seem to cluster around the most effective models for each scenario - where being able to accurately distinguish systems is critical. Moreover, we observe that there is a pattern to the occurring swaps - Triple2vec appears particularly favored under leave one last item, while VBCAR ranks much higher under temporal evaluation. This behaviour is likely being caused by both how the instances are being selected (e.g. whether evidence from the future is available when training) and the differing train/validation/test distributions (see Table~\ref{tab:dataset}). Hence, this provides evidence both that the splitting strategy is an important factor that impacts reported recommendation performance, and that a splitting strategy may favour particular systems.

\begin{table*}[tb]
	\caption{Performance comparison of recommendation models under different data splitting strategies. Models are sorted by performance under leave one last (item) splitting, arrows indicate rank position swaps relative to that performance.}
	\vspace{-3mm}
	\label{tab:swaps}
	\resizebox{150mm}{!}{
		\begin{tabular}{|l|ccc|c|l|ccc|}
			\cline{1-4}\cline{6-9}
			\multirow{2}{*}{Model} & \multicolumn{3}{c|}{Tafeng Dataset, NDCG@10} & & \multirow{2}{*}{Model} &  \multicolumn{3}{c|}{Tafeng Dataset, Recall@10} \\
			 & Leave One Item & Leave One Basket & Temporal Global & &  & Leave One Item & Leave One Basket & Temporal Global \\
			\cline{1-4}\cline{6-9}
		    NMF & 0.0879 & 0.0796 & 0.1811 & & NMF & 0.1739 & 0.0969 & 0.1671 \\
		    BPR & 0.1347 & 0.1987 & 0.2575 & & BPR & 0.2470 & 0.2306 & 0.2338\\
		    VAECF & 0.1580 & 0.2309 & 0.2858 & & VBCAR & 0.2835 & 0.2633$\blacktriangledown$(1) & 0.3129$\blacktriangledown$(3)\\
		    NeuMF & 0.1738 & 0.2504 & 0.3313 & & VAECF & 0.2861 & 0.2651$\blacktriangledown$(1) & 0.2655$\blacktriangle$(1)\\
		    VBCAR & 0.1739 & 0.2549 & 0.3744$\blacktriangledown$(1) & &  Triple2Vec & 0.2957 & 0.2622$\blacktriangle$(2) & 0.3055$\blacktriangle$(1) \\
		    NGCF & 0.1852 & 0.2726$\blacktriangledown$(1) &  0.3794$\blacktriangledown$(1) & & NeuMF & 0.3125 & 0.2881 & 0.3110$\blacktriangle$(1) \\
		    Triple2Vec & 0.1978 & 0.2555 $\blacktriangle$(1) & 0.3569$\blacktriangle$(2) & & NGCF & 0.3364 & 0.3112 & 0.3655 \\
		    \cline{1-4}\cline{6-9}
		    \multicolumn{9}{c}{}\\
		    \cline{1-4}\cline{6-9}
			\multirow{2}{*}{Model} & \multicolumn{3}{c|}{\zqc{Dunnhumby} Dataset, NDCG@10} & & \multirow{2}{*}{Model} &  \multicolumn{3}{c|}{\zqc{Dunnhumby} Dataset, Recall@10} \\
			 & Leave One Item & Leave One Basket & Temporal Global & & & Leave One Item & Leave One Basket & Temporal Global \\
			\cline{1-4}\cline{6-9}
		     BPR & 0.1354 & 0.2413  & 0.5264$\blacktriangledown$(1)  & & NMF & 0.2498 & 0.2514$\blacktriangledown$(3) & 0.0908 \\
		     NMF & 0.1448 & 0.2496  & 0.4327$\blacktriangle$(1)  & & BPR & 0.2514 & 0.2163$\blacktriangle$(1) & 0.1018 \\
		     VAECF & 0.1455 & 0.2620 & 0.5790  & & VAECF & 0.2759 & 0.2371$\blacktriangle$(1)  & 0.1173\\
		     NGCF & 0.1480 & 0.2647 & 0.6031 & & NGCF & 0.2836 & 0.2434$\blacktriangle$(1) & 0.1275 \\
		     NeuMF & 0.2080 & 0.3407  & 0.6376 & &  VBCAR & 0.3797 & 0.3342$\blacktriangledown$(2) & 0.1431$\blacktriangledown$(1) \\
		     VBCAR & 0.2518 & 0.3873$\blacktriangledown$(1) & 0.6804$\blacktriangledown$(1)  & & NeuMF & 0.3906 & 0.3220 & 0.1410$\blacktriangle$(1) \\
		     Triple2Vec & 0.3043 & 0.3607$\blacktriangle$(1) & 0.6761$\blacktriangle$(1) & & Triple2Vec  & 0.4391 & 0.3193$\blacktriangle$(2) & 0.1454 \\
		    \cline{1-4}\cline{6-9}
		\end{tabular}}
		\vspace{-3mm}
\end{table*}
% \todo{Kendall Tau --> Kendall's $\tau$ in figure 2}
% \todo{does this make a statement that one system needs more training data for each user than another system. I.e you need to post an intuition about why this is happening.}
\begin{figure}[t!]
	\centering
	\includegraphics[width = 120mm]{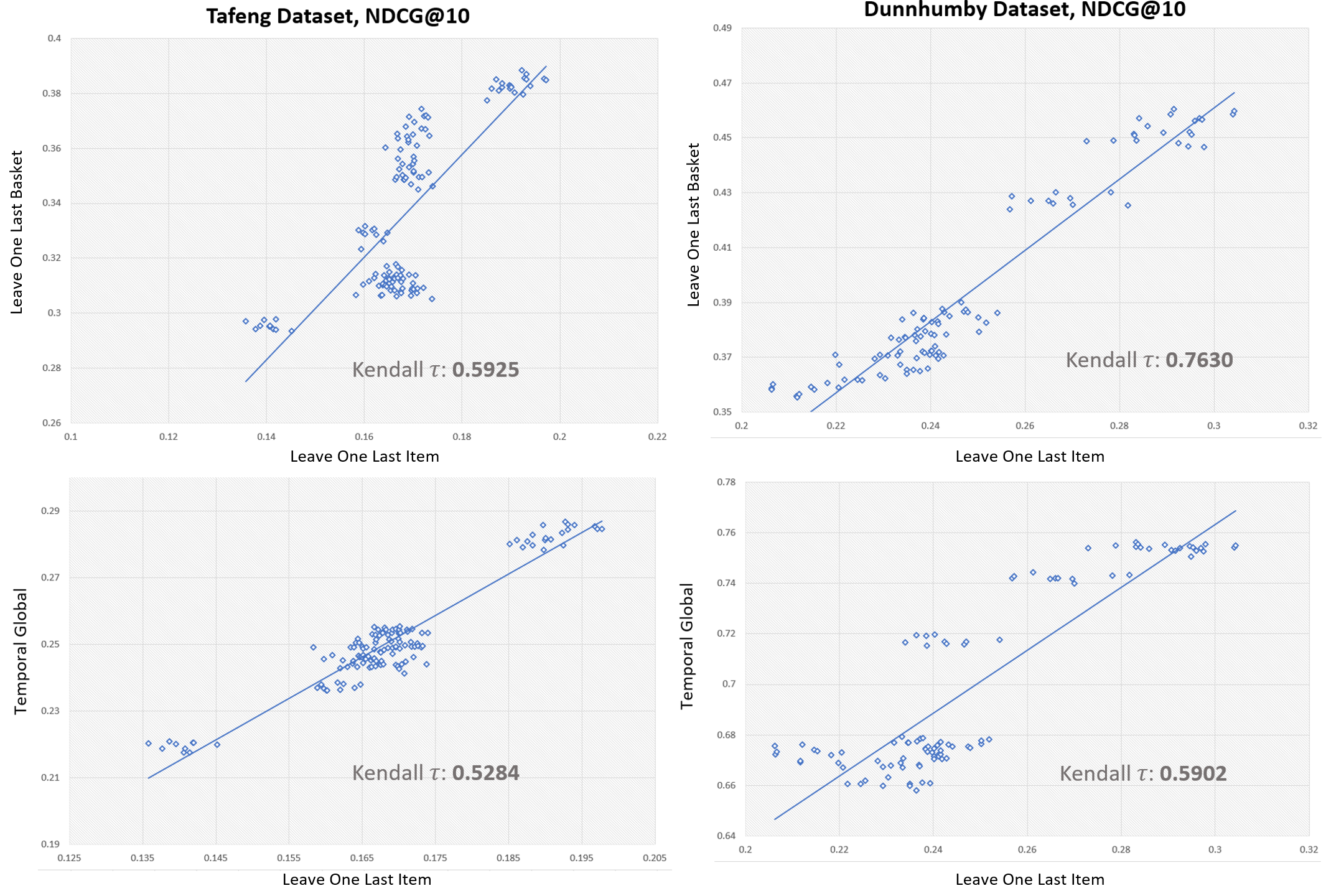}
	\vspace{-3mm}
	\caption{\label{fig:scatter}Splitting strategy pair-wise comparison under recommendation NDCG@10 for 230 models.}
	\vspace{-3mm}
\end{figure}

\enlargethispage{1\baselineskip}
\looseness -1 However, so far we have only considered 7 recommendation systems. With such a small sample size, these observed swaps could have simply occurred by chance. Hence, to test this, we perform a correlation experiment between a much larger sample of recommendation systems. In particular, we take three of the more effective learning algorithms (NeuMF, VBCAR and Triple2Vec), and generated 230 models by varying their hyperparameters, providing a larger sample set to compare. Figure~\ref{fig:scatter} plots the NDCG@10 performance of these models for pairs of splitting strategies across each of the two datasets, as well as reporting Kendall's $\tau$ correlation between the score distributions for each. If a pair of splitting strategies produces a similar ranking of systems, we would expect a Kendall's $\tau$ value close to 1.0 and the data points to align close to the linear trend line. As we can see from Figure~\ref{fig:scatter}, the Kendall's $\tau$ correlations between the pairs of splitting strategies are only moderate, ranging between 0.5284 and 0.7630, demonstrating that there are many rank swaps taking place. Additionally, we can see that at the higher end of the effectiveness scale (top-right of each chart), there is greater horizontal point dispersion than vertical point dispersion. This means that leave one last item data splitting is producing a wider distribution of NDCG@10 and Recall@10 scores, while the temporal and leave one last basket splitting seems to group systems in a more stratified manner. This does not indicate that one strategy is better than another, but is evidence that these splitting strategies are in effect evaluating very different aspects of recommendation.

\looseness -1 To conclude, we have shown that the ranking of state-of-the-art systems is strongly affected by the data splitting strategy employed, and hence, is a confounding variable that needs to be accounted for when comparing recommendation systems. We have also observed some evidence that certain splitting strategies may favour particular systems.\footnote{This is an important direction for future work.}

\begin{comment}
\begin{figure}[t!]
	\centering
	\includegraphics[width = 95mm]{Kendall.pdf}
	\caption{\label{fig:data_split}Kendall's $\tau$ of the performance ranking over the three data split strategies, where L1B, L1I and TEM denote the leave\_one\_basket\_out, leave\_one\_item\_out and global temporal data split strategies respectively.}
\end{figure}
\end{comment}

\enlargethispage{1\baselineskip}
\section{Conclusions}
\looseness -2 In this work, we analyzed the impact that different data splitting strategies have on the reported performance of different recommendation models, as the splitting strategies used in the literature vary greatly. Through experimentation \zqc{using} three splitting strategies, seven recommendation models and two datasets, we have shown that the splitting strategy employed is an important confounding variable that can markedly alter the ranking of state-of-the-art systems. This is important, as it highlights that much of the current research being published is not directly comparable, even when the same dataset and metrics are being used. Furthermore, we also have observed that certain splitting strategies favour particular recommendation models - potentially due the different balance of train/validation/test data under each scenario and factors such as whether future evidence is available during training.
% \todo{repeat conclusion about whether some model needs more training data or something}
In terms of best practices for future researchers, we recommend the following:

\vspace{1mm}
\looseness -1 \noindent \textbf{1) Report the splitting strategy employed}: This includes the statistics of the train/validation/test components and any user/item filtering performed, as these can strongly impact performance.

\noindent \textbf{2) \zqc{Report} performance under temporal global splitting}: This is generally seen as the most realistic setting, and so should be the default splitting strategy used.

\noindent \textbf{3) Release your data splits}: so that they can be re-used by other researchers. The data splits used in this work can be downloaded from \textbf{\url{http://github.com/anonymdata/data_splits}}.
\enlargethispage{1\baselineskip}

\bibliography{bib}
\bibliographystyle{ACM-Reference-Format}
\end{document}